\documentclass[final,3p,times,twocolumn]{elsarticle}

\usepackage{lineno}

\usepackage{gensymb}

\journal{Nuclear Physics B}

\begin{document}

\begin{frontmatter}



\title{Characterization of self-injected electron beams
from LWFA experiments at SPARC\_LAB}

\author[infn]{G. Costa}
\author[infn]{M. P. Anania}
\author[infn]{F. Bisesto}
\author[infn]{E. Chiadroni}
\author[Tor Vergata]{A. Cianchi}
\author[Sapienza]{A. Curcio}
\author[infn]{M. Ferrario}
\author[infn]{F. Filippi}
\author[infn]{A. Marocchino}
\author[infn]{F. Mira}
\author[infn]{R. Pompili}
\author[Israel]{A. Zigler}

\address[infn]{Laboratori Nazionali di Frascati, INFN, Via E. Fermi, 40, 00044 Frascati, Italia}
\address[Tor Vergata]{University of Rome Tor Vergata, Via della Ricerca Scientifica, 1, 00133 Roma, Italia}
\address[Sapienza]{University of Rome La Sapienza, Piazzale Aldo Moro, 5, 00185 Roma, Italia}
\address[Israel]{Racah Institute of Physics, Hebrew University, 91904 Jerusalem, Israel}

\begin{abstract}
The plasma-based acceleration is an encouraging technique to overcome the limits of the accelerating gradient in the conventional RF acceleration. A plasma accelerator is able to provide accelerating fields up to hundreds of $GeV/m$, paving the way to accelerate particles to several $MeV$ over a short distance (below the millimetre range). 

Here the characteristics of preliminary electron beams obtained with the self-injection mechanism produced with the FLAME high-power laser at the SPARC\_LAB test facility are shown.

In detail, with an energy laser on focus of $1.5\ J$ and a pulse temporal length (FWHM) of $40\ fs$, we obtained an electron plasma density due to laser ionization of about $6\times10^{18}\ cm^{-3}$, electron energy up to $350\ MeV$ and beam charge in the range $(50-100)\ pC$.
\end{abstract}

\end{frontmatter}

\section{Introduction}

Particle accelerators are nowadays used in a wide range of fields: they are the fundamental tools for elementary particle research and high-energy physics in general, but at the same time they are an important resource employed in several other areas, from medicine to the industry \cite{eliezer2008applications}. 

The goal of the acceleration of charged particles has been to continually achieve higher energies and, due to limitations on the maximum electric fields that can be obtained by conventional techniques, ever larger accelerator systems have become necessary, so the cost of such systems is so huge that multinational collaboration and construction over many years is required. In fact, RF structures can sustain accelerating fields up to $100\ MV/m$, limited by breakdown, therefore, hundreds-of-metres-long machines are needed to achieve beams at $GeV$ scale.

In 1979, Tajima and Dawson proposed a new method to accelerate charged particles using laser-plasma interaction (Laser WakeField Acceleration, LWFA) \cite{tajima1979laser}. With this technique, a short laser pulse interacts with a plasma generating a strong oscillating electrostatic field in its wake. The novelty of this technique is that the electrostatic fields generated by such a process can be several orders of magnitude larger than that which can be generated by conventional radio frequency cavities in accelerators. Indeed, the plasma accelerating field scales as $E_0[V/m]\simeq 96 \sqrt{n_0[cm^{-3}]}$, where $n_0$ is the electron plasma density \cite{esarey2009physics}. Therefore, accelerating gradients up to $TeV/m$ are achievable. In the last ten years, thanks to the implementation of the Chirped Pulse Amplification (CPA) method \cite{strickland1985compression}, the development of $TW$-class $fs$-level lasers has been fundamental to study LWFA \cite{pukhov2004bubble} and electron beams with energy at GeV-scale have been experimentally produced \cite{geddes2004high,leemans2006gev,fuchs2009laser,mangles2004monoenergetic}. 

Our experiment about the characterization of preliminary electron beams, produced by self-injection at the SPARC\_LAB test facility \cite{ferrario2013sparc_lab} (INFN-LNF) by using the ultra-short high power laser FLAME \cite{bisesto2017flame}, is placed in this context. The latter is a CPA Ti:Sa laser system, a compact femtosecond laser source, providing up to $7.5\ J$ pulse energy at $10\ Hz$ repetition rate, with a bandwidth of $80\ nm$ and a contrast ratio in the order of $10^{10}$. The pulse can be as low as $25\ fs$ and leads to a peak power higher than $250\ TW$. The goal of our experimental work was to reproduce the strongly non-linear regime, therefore extremely relativistic laser intensities, testing our experimental set-up and confirming our specific simulation. 

\section{Experimental parameters and simulations}
\begin{figure}[!htbp]
	\centering
	\includegraphics[width=2.5in]{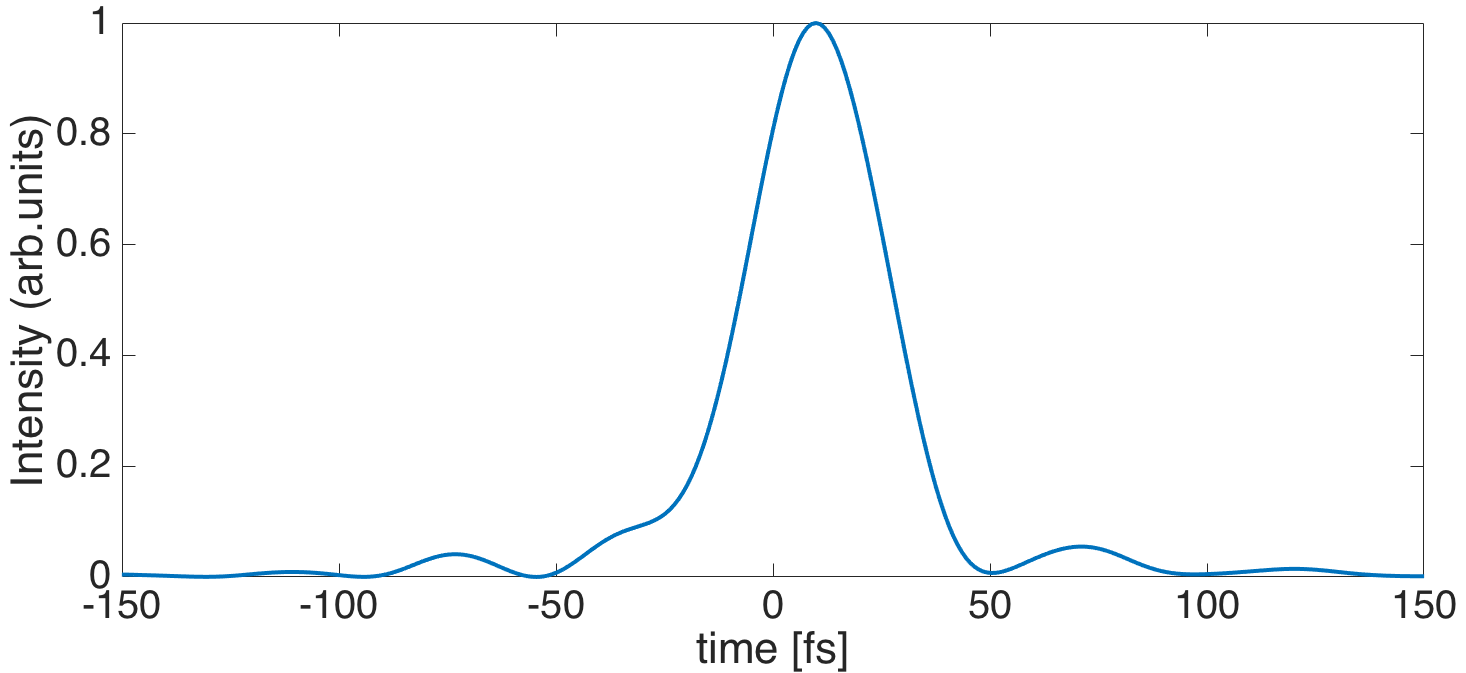}
	\caption{FLAME temporal pulse length measurement made using the APE SPIDER after the compression.}\label{spider}
\end{figure}

\begin{figure}[!htbp]
	\centering
	\includegraphics[width=1.3in]{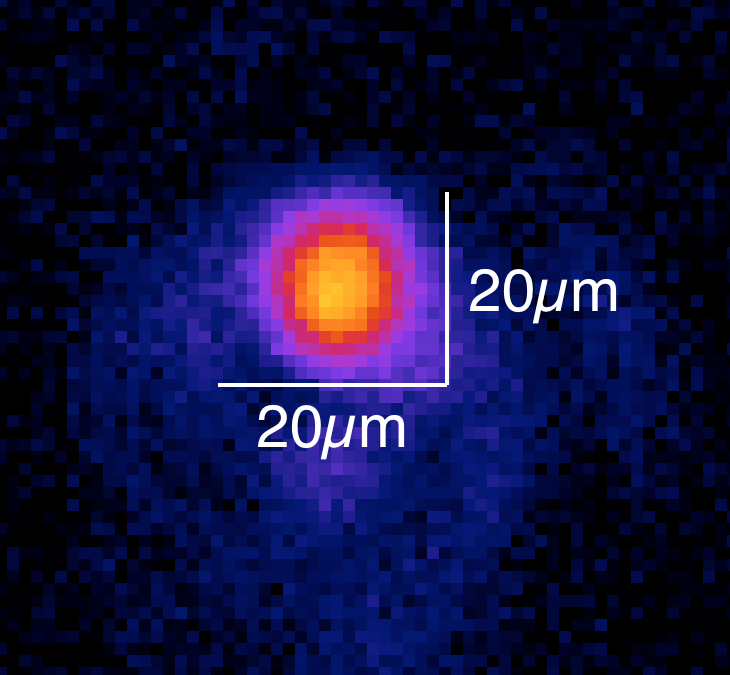}
	\caption{FLAME focal spot in vacuum at the gas-jet position, obtained by means of a Basler Scout scA640-70gm CCD camera equipped with a microscope objective.}\label{focus}
\end{figure}

To reproduce the strongly non-linear bubble regime, the main laser pulse parameters were fixed according to the scaling laws for laser wakefield accelerators \cite{pukhov2002laser,tajima2017laser}: the energy pulse was within the values $E = (1-1.5)\ J$, delivered in $\tau_L = (35-40)\ fs$ temporal length (see Fig.\ref{spider}), over a $10\ \mu m/e^2$ radius on focus (see Fig. \ref{focus}). In this way we had a laser intensity $I_0 = 2 E_0/\tau_L \pi r^2 \approx 2 \times 10^{19}\ W/cm^2$, therefore the laser normalized intensity \cite{esarey2009physics,pukhov2002laser} is $a_0 \simeq 0.855 \times 10^{-9} \lambda_L \sqrt{I_0} \approx 3.3$, confirming the strongly non-linear regime. 

To reach the maximum electron beam energy, given by $E_b = 2 m_e c^2 a_0 n_c/3 n_e$, where $n_c = \pi/r_e \lambda_L^2$ is the critical plasma density, $r_e$ is the Rayleigh length and $\lambda_L =800\ nm$ is the laser wavelength, it is necessary to set the acceleration length $L_{acc}$ to the dephasing length: $ L_{acc} = L_{deph} = (2/3\pi)\sqrt{a_o \lambda_L} (n_c/n_e)^{3/2}$, while the pump depletion length $L_{depl} \simeq c \tau_L (n_c/n_e) \geq L_{deph}$. In this way, for a given $a_0$, a longer pump depletion length, corresponding to a longer dephasing length, is due to an increase of the critical density, which indicates a decrease of the electron self-injection threshold into the laser wakefield as well as a decrease of the critical power for the self-guiding \cite{tajima2017laser}.

\begin{figure}[!htbp]
 	\centering
 	\includegraphics[width=1.9in]{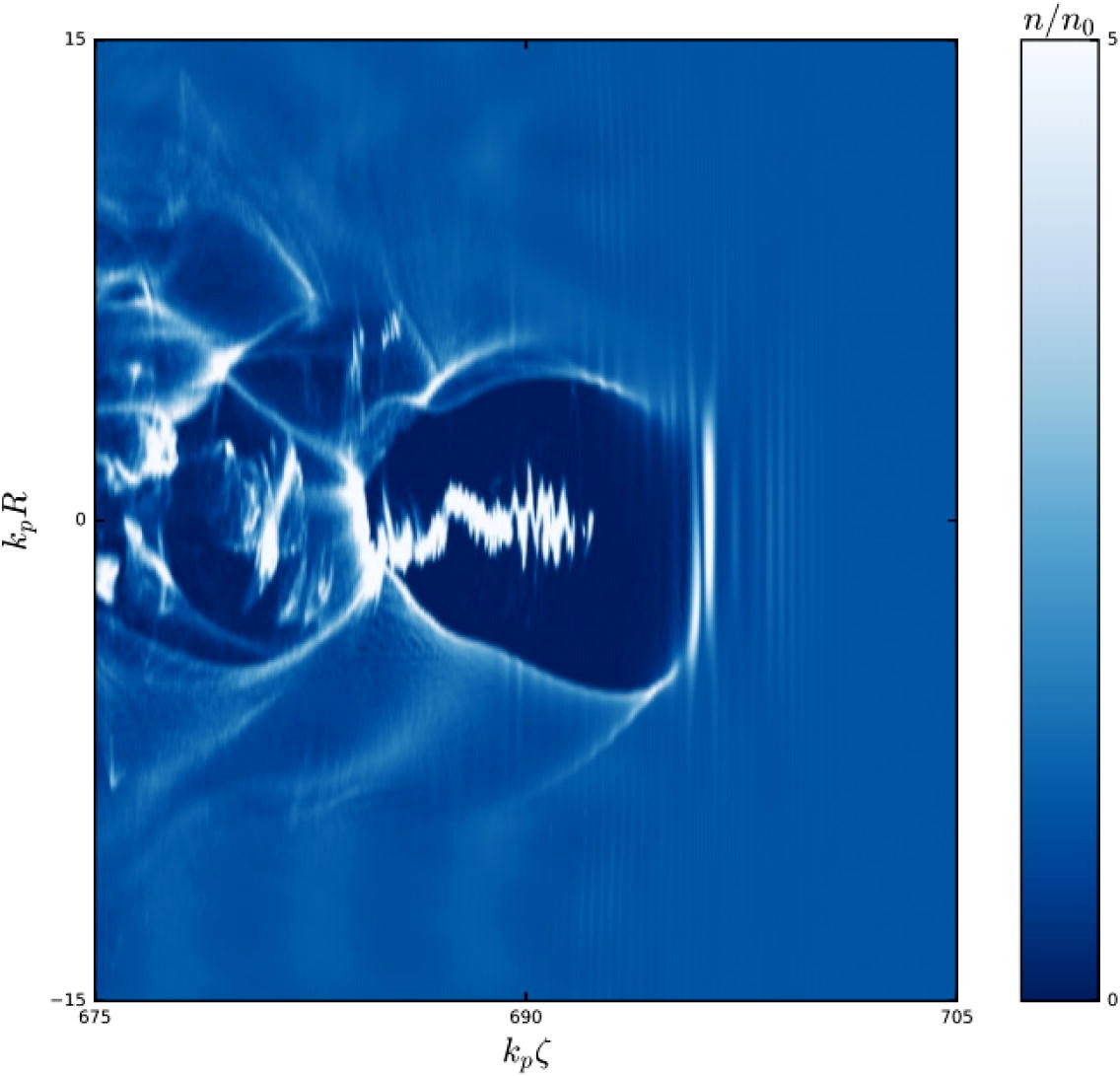}
 	\caption{Example of a 2D plasma density plot from PIC Code ALaDyn simulations, in dimensionless units. The snapshot is taken at $0.7\ mm$ propagation distance, the laser FWHM duration is $30\ fs$, while the initial laser diameter is $11.76\ \mu m$. For the full 3D simulation it was used a $1120 \times 304 \times 304$ moving window, with a longitudinal resolution of $0.05\ \mu m$, a transverse resolution of $0.0125\ \mu m$ and 8 particles per cell. The plasma density is normalized to the unperturbed background density ($n_e = 1 \times 10^{19}\ cm^{-3}$)  \cite{curcio2017first}.}\label{aladyn}
\end{figure}

Our regime of interaction was confirmed also by full 3D PIC simulations (see Fig. \ref{aladyn}), performed with the ALaDyn code \cite{benedetti2008tt,londrillo2010charge,marocchino2017aladyn} for a driver laser pulse and a background electron plasma density with about the same characteristics of the experimental ones. The code predicted polyenergetic electron beams, with energies and energy spreads both consistent with the measurements \cite{curcio2017first}.

\section{Experimental set-up and results}
\begin{figure}[!htbp]
 	\centering
 	\includegraphics[width=1.7in]{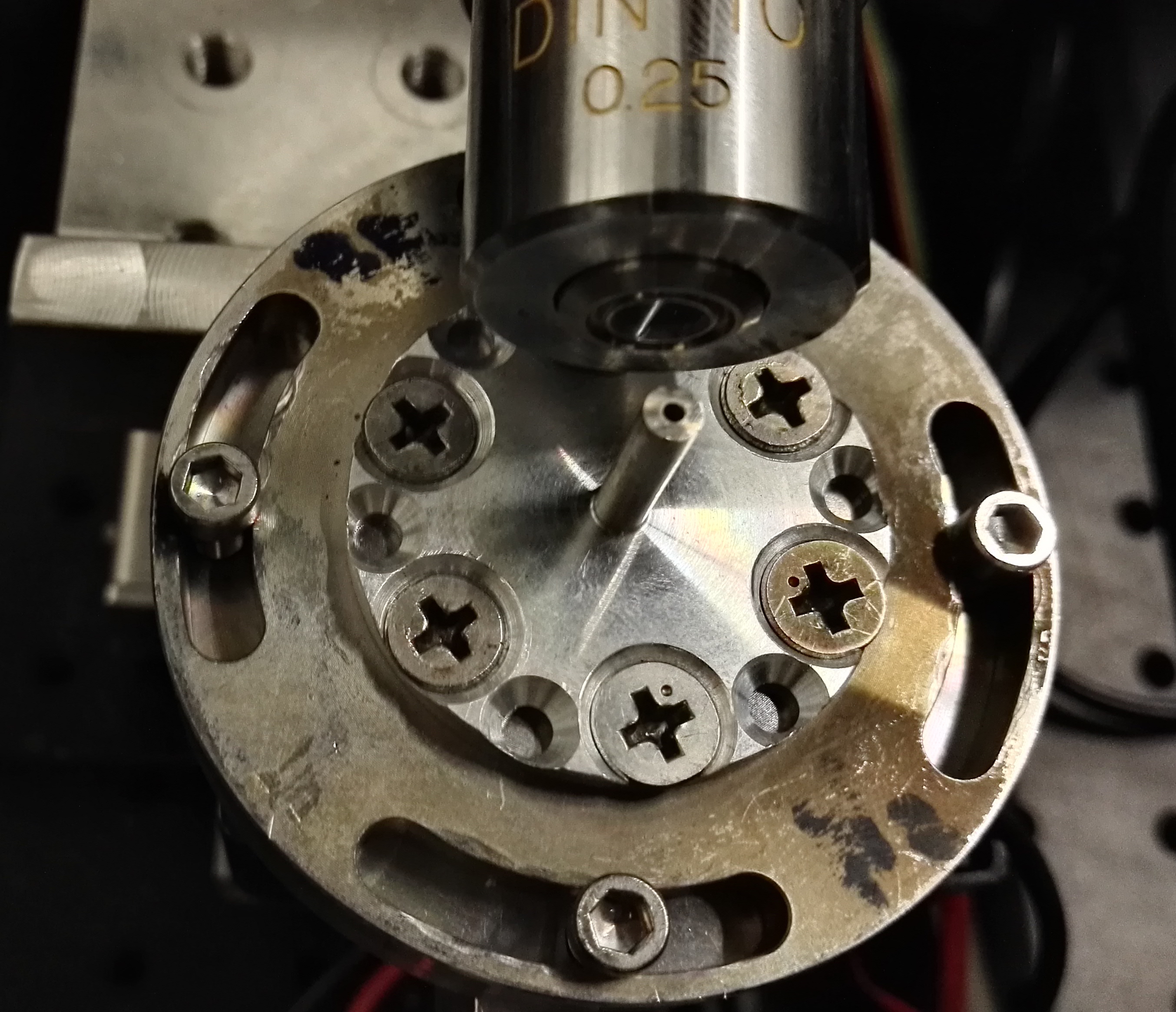}
 	\caption{FLAME supersonic gas-jet, the design of the nozzle allows to obtain a gas column with a sharp edge.}\label{gasjet}
\end{figure}

The laser was focused by means of an off-axis parabolic (OAP) mirror with $1\ m$ focal length in the middle of the vacuum ($10^{-6}\ mbar$) interaction chamber on a specific $He$ gas-jet target. Outside the vacuum chamber was placed a gas bottle equipped with a specific knob used to throw gas with variable pressure ($0-50)\ bar$. The gas-jet nozzle, displayed in Fig. \ref{gasjet}, has been designed to provide a supersonic gas profile, which obtained a sharp boundary with the vacuum. The flow of gas into the vacuum was optimized and minimized using a solenoid valve that was normally closed, the gas was injected synchronously with the laser pulse so that when the laser arrived into the chamber, it could ionize the gas and bring it into the state of plasma.

\begin{figure}[!htbp]
 	\centering
 	\includegraphics[width=2.7in]{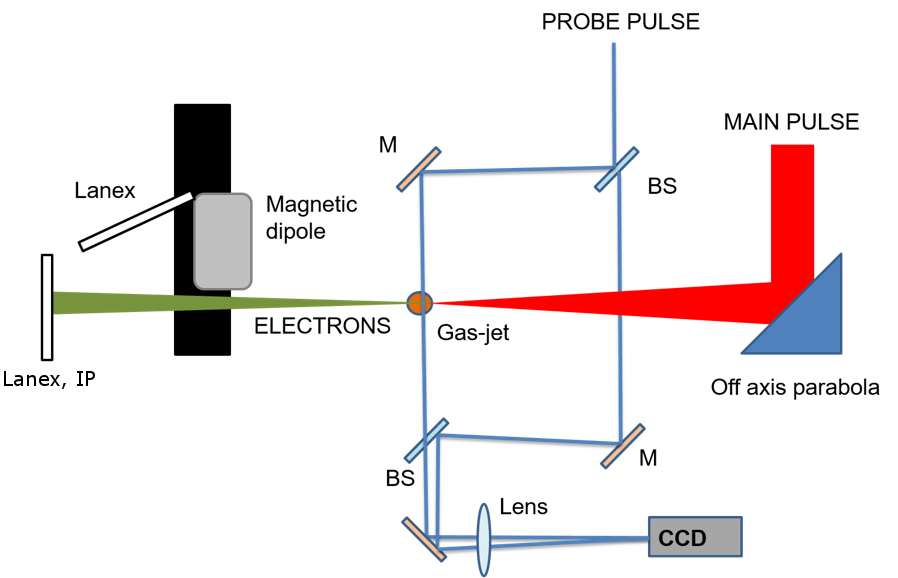}
 	\caption{Self-injection experimental set-up composed by a Mach-Zehnder interferometer for plasma density measurement (where BS: Beam Splitter, M: Mirror), a magnetic spectrometer for electron energy measurement, a LANEX screen and an IP for electron bunch transversal dimensions and charge.}\label{setup}
\end{figure}

The experimental set-up, shown in Fig. \ref{setup}, was composed by a Mach-Zehnder interferometer to measure the generated plasma density, a LANEX screen scintillator to detect the electron bunch transverse dimensions, a Fuji BAS Imaging Plate (IP) with a specific Durr CR35 Bio scanner to evaluate the electron bunch charge, and a $1\ T$ permanent dipole coupled to another LANEX screen scintillator oriented at $60\degree$ with respect to the direction of the beam propagation to obtain the electron energy spectrum.

\subsection{Plasma density and acceleration length}
\begin{figure}[!htbp]
  	\centering
  	\includegraphics[width=2in]{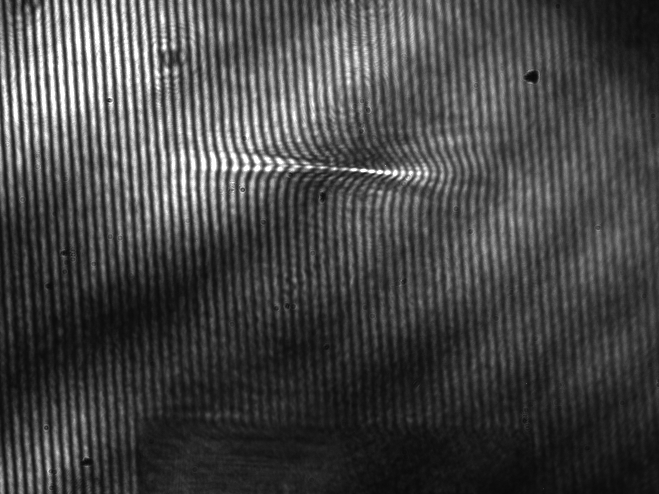}
  	\caption{Typical interferogram in the presence of plasma. The spatial calibration has been performed by using the shadow of the gas-jet, whose thickness is known.}\label{interferogram}
\end{figure}

The Mach-Zehnder interferometer uses the dependence of the plasma refraction index $\eta = \sqrt{1-n_e/n_c}$ on the medium density, and its variation can be obtained measuring the phase shifting of a propagating beam. For this purpose, we used a probe beam ($10\%$ fraction of the main beam), that was split in two arms, while only one was passing through the plasma. Observing their recombination, we were able to look at the image of the plasma channel through a $25\ cm$ focal length lens, while the interferogram was collected by a Basler Scout 750 CCD camera (see Fig. \ref{interferogram}). The dephasing of the interference fringes, due to the transit through the plasma, was analysed with a specific Matlab GUI program, which reconstructs the phase map obtained by the shift of the fringes recorded \cite{filippi2017thesis}.  

The electron plasma density obtained was $n_e \approx (6-8)\times 10^{18}\ cm^{-3}$, and the maximum acceleration length, measured looking at the extension of the plasma in the interferometric images of the laser propagated through the plasma, was in the range $L_{acc} = (1-2)\ mm$, according to the diameter of the gas-nozzle and of the plasma density used shot by shot \cite{curcio2017first}. 

Using these experimental measurements, it was possible to calculate the maximum electric field reached in the plasma, that was $E_0 \approx 240\ GV/m$, considering a plasma density equal to $6\times 10^{18}\ cm^{-3}$, value confirmed by the fact that the bubble radius is $R_{bubble} = 2\sqrt{a_0}c/\omega_p \approx \lambda_p/2$, where $\omega_p \simeq e\sqrt{n_0/(\epsilon_0 m_e)}$ is the plasma frequency and $\lambda_p = 2\pi c/\omega_p$ \cite{esarey2009physics}. 

In this way, for our parameters, we could notice that for an acceleration length, measured as the plasma channel length, as low as $L_{acc} = 1.5\ mm$, we obtained $L_{acc}<L_{deph}\approx 1.55\ mm<L_{depl}\approx 3.5\ mm$, and reached the maximum energy gain $E_b \approx 330\ MeV$ for our system. 

The channel length was also measured by images recorded using $90\degree$ Thomson scattering diagnostic, which provided on-line data of the electron-photon collisions, due to the interaction between plasma and the laser pulse, used also to check the best conditions for electron acceleration. It was set up with a $200\ mm$ focal length lens, which provided the possibility to get, on the CCD camera (Basler scA1600-14gc), a magnified image of the region of interest, and a $35\ mm$ objective \cite{bisesto2017thesis}. 

\subsection{Electron bunch dimensions, charge and energy}
\begin{figure}[!htbp]
 	\centering
 	\includegraphics[width=1.7in]{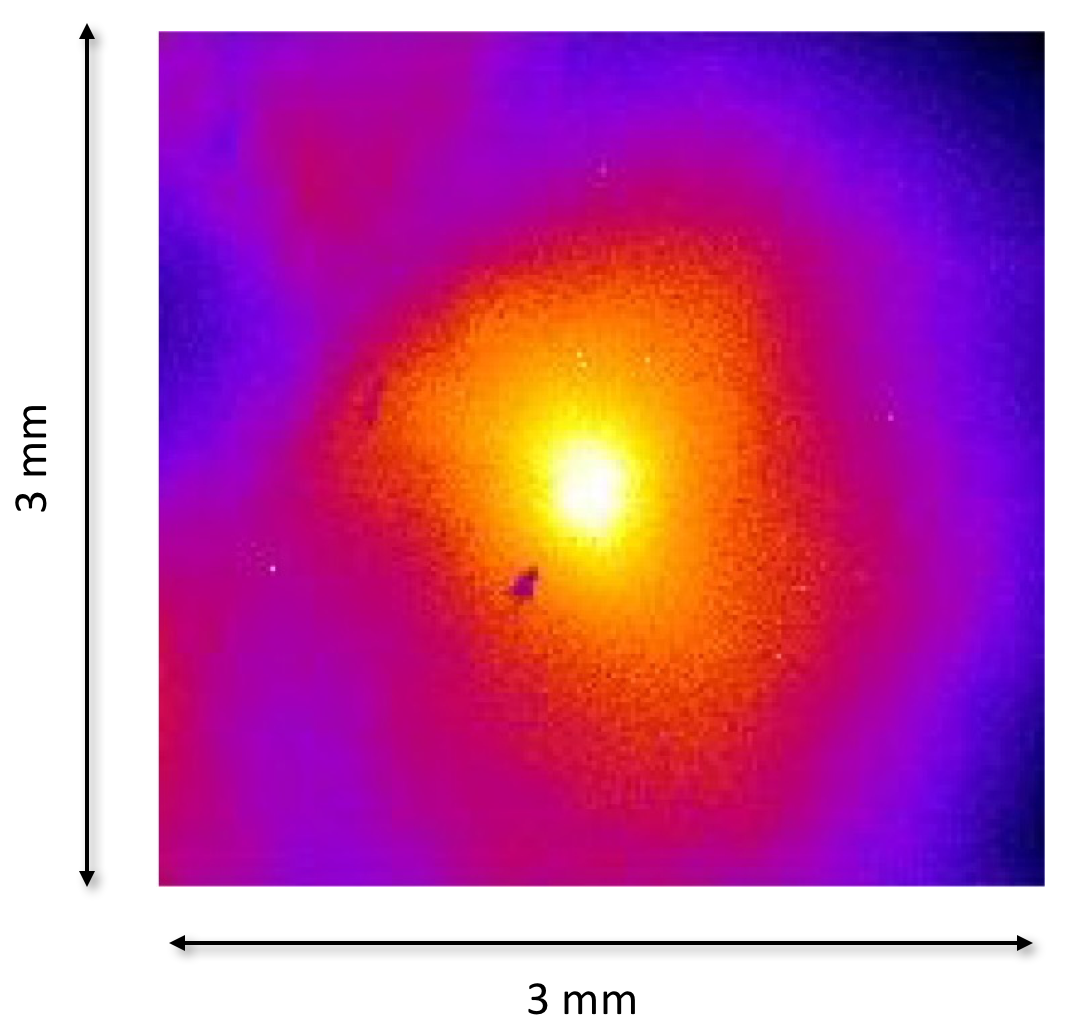}%
 	\\\includegraphics[width=3in]{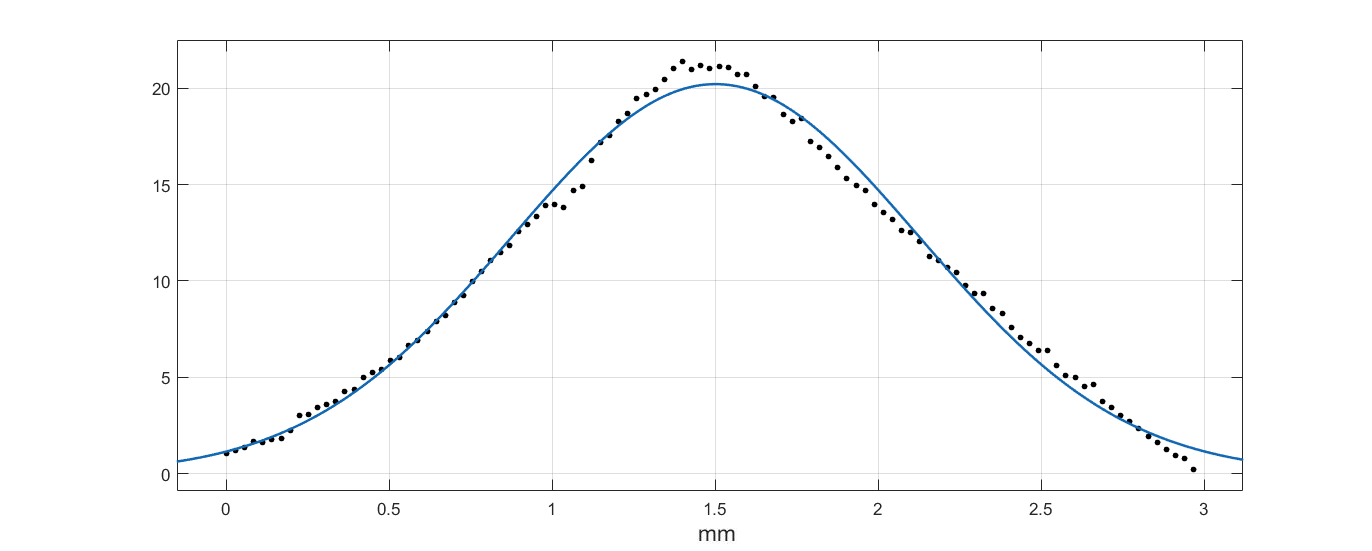}
 	\caption{Typical measurement of electron beam transverse profile. Axes are Intensity [arb. units] vs dimension [mm]; data (black dots), fit (blue curve).}\label{lanex_profile}
\end{figure}

Measurements of electron beam transverse profile were performed with the LANEX screen, placed $50\ cm$ far from the source, outside the experimental chamber at $90\degree$ with respect to the direction of the electrons, i.e. the propagation direction of the laser pulse. A CCD camera (Basler scA640-70gm), looking at the screen, has been used to image it. A typical profile is in Fig. \ref{lanex_profile}: the dimension of the obtained electron spot was in the range $(1-3)\ mm$ FWHM, which corresponds to a divergence in the range $(2-6)\ mrad$. 

\begin{figure}[!htbp]
 	\centering
 	\includegraphics[width=1.7in]{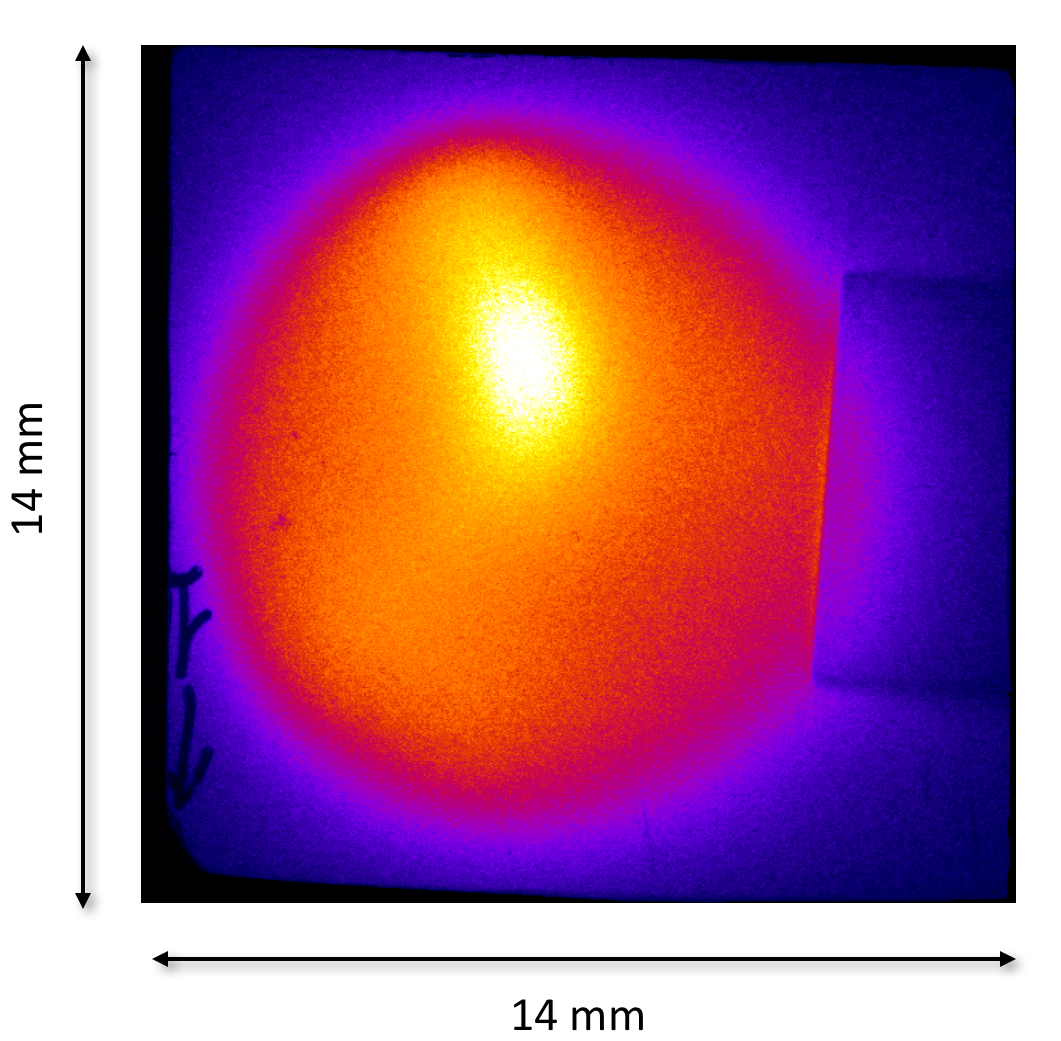}%
 	\\\includegraphics[width=3in, height=1in]{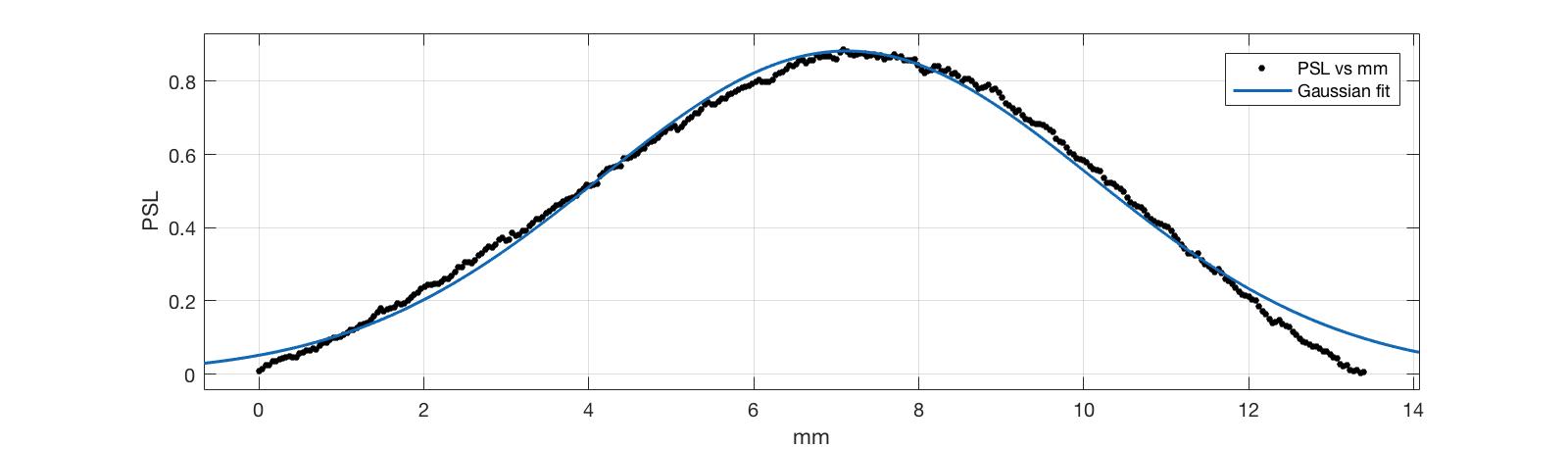}
 	\caption{Typical measurement of IP spot and converted integrated profile. Axes are PSL vs dimension [mm]; data (black dots), fit (blue curve).}\label{ip_profile}
\end{figure}

By replacing this LANEX screen with an IP and reading it with our Durr scanner, we were able to retrieve the deposited electron radiation through the Gray Value data shown in Fig. \ref{ip_profile}. It was then converted in PSL (photostimulated luminescence) via the calibration presented in the work \cite{curcio2017first,curcio2016imaging,nakanii2008absolute}, leading us to an evaluation of the charge typically in the range $(50-100)\ pC$. The charge data confirmed the electron beam transversal information found with the LANEX screen.

\begin{figure}[!htbp]
 	\centering
 	\includegraphics[width=1.7in]{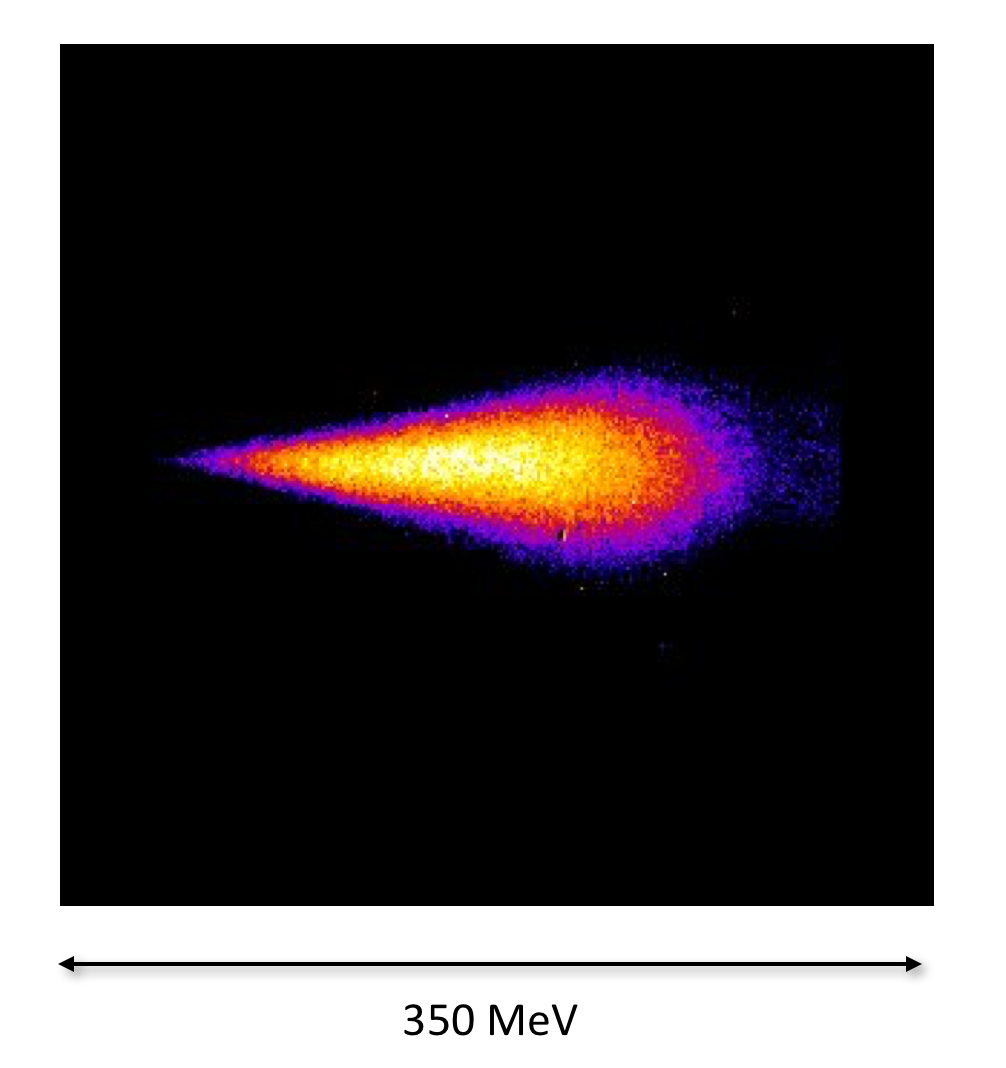}%
 	\\\includegraphics[width=3in]{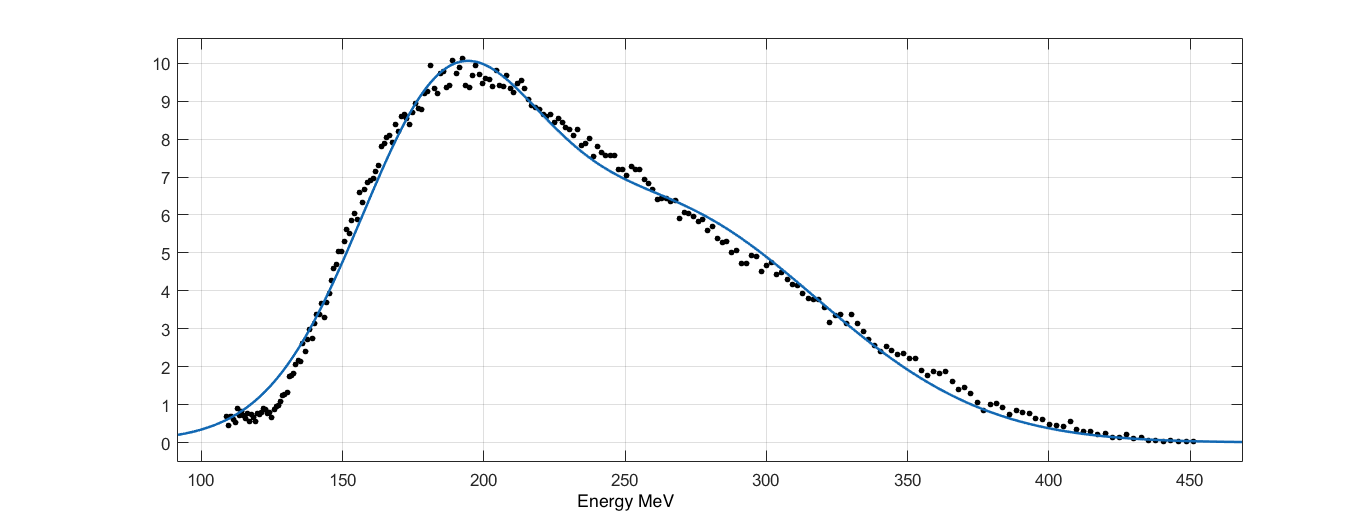}
 	\caption{Typical electron energy spectrum measurement. Axes are Intensity [arb. units] vs Energy [MeV]; data (black dots), fit (blue curve).}\label{energy_profile}
\end{figure}

Simulations made with the General Particle Tracer (GPT) software \cite{de1996general} were used to optimize the magnetic spectrometer setup in order to measure the electron energy and to find the right correlation between screen position and electron energy \cite{bisesto2017thesis}. The values obtained were in the range $(150 - 350)\ MeV$ (Fig. \ref{energy_profile}), with an energy spread of about $(10-15)\%$, therefore measurements consistent with the expected maximum energy gain $E_b$.

\section{Conclusions and future developments}
Experimental results about the characterization of preliminary electron beams, produced by the self-injection bubble regime at SPARC\_LAB have been shown. Electrons produced with the FLAME laser were mainly used to perform diagnostic tests, based on betatron \cite{curcio2017trace,curcio2017single} and optical transition radiation \cite{bisesto2017novel,bisesto2017innovative,bisesto2017innovatives}.

In the near future, we are developing a magnetic transport line for the electrons coming from the plasma source, in order to better control them. In particular, we are designing a particle selector to keep only the energetic core of the beam and a magnetic triplet to control the electron divergence, typically high in a plasma accelerator, and tune the bunch transverse beam size. Moreover, an integrated current transformer (ICT) will be installed in the target area to measure the electron charge. On the other side, an optimized injection method will be studied to reduce the energy spread. In particular, an experiment based on external injection scheme is foreseen, combining the high brightness SPARC\_LAB photoinjector and the FLAME laser \cite{bisesto2016laser,rossi2016stability,rossi2014external}.  

\section*{Acknowledgements}
This work was supported by the European Union's Horizon 2020 research and innovation programme under grant agreement No. 653782.

\section*{References}
\nocite{*}
\bibliography{bibliography}

\end{document}